\documentclass[twocolumn,showpacs,preprintnumbers,amsmath,amssymb]{revtex4}

\usepackage[dvips]{epsfig}
\usepackage{graphicx}
\newcommand{\be}{\begin{equation}}
\newcommand{\ee}{\end{equation}}
\newcommand{\bea}{\begin{eqnarray}}
\newcommand{\eea}{\end{eqnarray}}

\begin{document}

\title{Phase Behavior and Selectivity of DNA-linked Nanoparticle
Assemblies}
\author{D. B. Lukatsky and Daan Frenkel}
\affiliation{FOM Institute for Atomic and Molecular Physics,
Kruislaan 407, 1098 SJ Amsterdam, The Netherlands}
\date{\today}

\begin{abstract}
We propose a model that can account for the experimentally
observed phase behavior of DNA-nanoparticle assemblies (R. Jin et
al., JACS \textbf{125}, 1643 (2003); T. A. Taton et al., Science
\textbf{289}, 1757 (2000)). The binding of DNA-coated
nano-particles by dissolved DNA linker can be described by
exploiting an analogy with quantum particles obeying fractional
statistics. In accordance with experimental findings, we predict
that the phase-separation temperature of the nano-colloids
increases with the DNA coverage of the colloidal surface. Upon the
addition of salt, the demixing temperature increases
logarithmically with the salt concentration. Our analysis suggests
 an experimental strategy to map
\textit{microscopic} DNA sequences onto the \textit{macroscopic}
phase behavior of the DNA-nanoparticle solutions. Such an approach
should enhance the efficiency of methods to detect (single)
mutations in specific DNA sequences.
\end{abstract}
\maketitle

Colloidal particles that can be selectively linked by specific
sequences single-stranded DNA, represent an entirely novel class
of complex liquids. With these particles, it becomes feasible to
``design'' multicomponent mixtures, where the attractive
interaction between every pair of species can be controlled
independently.

\begin{figure}[ht]
\begin{center} 
\centerline{\psfig{figure=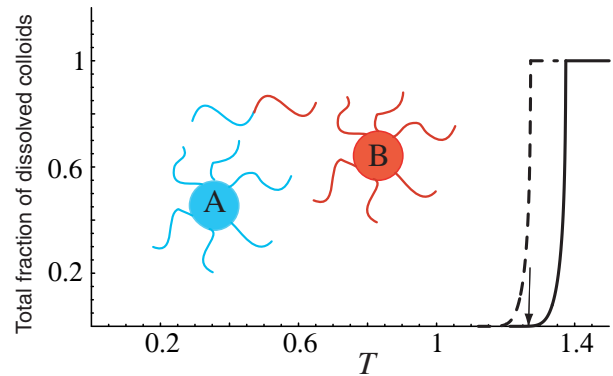,height=5cm,width=8cm}}
\caption{Computed temperature dependence of the  fraction
of the ssDNA-coated colloids in the fluid (dilute) phase of a
binary colloidal mixture (species $A$ and $B$) and added ssDNA
linkers. Above the demixing temperature, the volume fractions of
both the $A$ and $B$ colloids are equal to 20\%. The target ssDNA
can form links between $A$ and $B$ colloids, but not between $A$ and $A$,
or $B$ and $B$. At low temperatures, virtually all colloids end up in
the dense ``gel'' phase.
The two curves illustrate the effect of a  mutation in the target
DNA. The solid  and dashed curves were obtained with
$\mu\equiv\mu_{AB}=3.5$, $s=10$, $q=6$, $\overline{M}=5$. All
temperatures are relative to the melting temperature of the
perfectly complementary linker (solid curve). The dsDNA melting
temperature for the dashed curve (corresponding to a base-pair
mismatch) is 10 \% lower than $T_m$ for the perfectly matched
linker. The arrow indicates the optimal
stringency temperature.}
\label{figAB}
\end{center}
\end{figure}

During the past few years, the unique molecular recognition
properties of DNA have been exploited to create self-assembled
nano-structures and nano-devices with a remarkable degree of
control (see e.g. Ref.~\cite{Seeman03}). Self-assembled
nanostructures based on DNA have also been utilized to detect
specific DNA sequences~\cite{Mirkin96,Mirkin00,Mirkin03,Kiang03}. In
nanoparticle-based DNA detection systems, single-stranded DNA
(ssDNA) ``probe" molecules are chemisorbed onto the surface of
gold nanoparticles (Fig.~\ref{figAB}). These probe ssDNA's can
specifically bind to a dissolved ``target" ssDNA  and thus detect
its presence with high sensitivity and selectivity: at high enough
concentration, the target ssDNA strands induces a sharp demixing
transition~\cite{fnmelting} that leads to  aggregation of the
nanoparticles coated with the complementary ``probe'' strands.
These aggregates can be easily detected by optical means.

The sensitivity of the nanoparticle-based DNA detection method is
some two orders of magnitude higher than that of the corresponding
fluorophore-based DNA-array scheme~\cite{Mirkin00}. But, in
addition, the method can be used to distinguish different DNA
sequences that differ from each other by only a single base. The
reason is that a single mismatch in the DNA sequence results in a
significant change in the DNA-colloid demixing temperature. The
fluorophore-based systems lack this level of selectivity
\cite{Mirkin00}.

In a recent paper, Jin et al.~\cite{Mirkin03} have analyzed the
dependence of the dissolution temperature (see
note~\cite{fnmelting}) of DNA-linked nanoparticles on the DNA
coverage density, the particle size, and the salt concentration.
In the experiments~\cite{Mirkin00,Mirkin03} short DNA molecules
($\sim20$-base oligonucleotides) and small gold colloids ($\sim10-50$ nm) were used.
The principal experimental findings are the following: (i) The
higher the ssDNA coverage density on the surface of colloids, the
sharper the dissolution profiles, and the higher the dissolution
temperature. (ii) The addition of salt stabilizes the aggregates.
The dissolution temperature increases logarithmically with salt
concentration. The higher the salt concentration, the larger the
total fraction of DNA-nanoparticle aggregates.

Below, we propose an explanation for these experimental findings.
We argue that the origin of the sharp phase-transition profiles is
the \textit{entropic cooperativity} of the DNA-nanoparticle
network. Upon cooling, the system undergoes liquid-liquid phase
separation. The dense ``liquid'' phase is strongly cross-linked
and behaves as a solid gel.

We first consider a simple model for a binary mixture of
DNA-coated colloids. Each colloidal species is covered with
specific ssDNA molecules, in such a way that all possible pairs of
colloids can be bound by three types of complementary ssDNA linkers:
\textit{viz.}, $AA$, $BB$, and $AB$.

For the sake of simplicity, we use a lattice model to analyze the
phase behavior of this system. Colloids of type $A$ and $B$ can
occupy the  nodes of a three-dimensional lattice with
coordination number $q$.  The system is grand canonical with
respect to all the components: $\mu^c_{A}$ and $\mu^c_{B}$ are the
chemical potentials of colloids of species $A$ and $B$,
respectively; and $\mu_{AA}$, $\mu_{BB}$, and $\mu_{AB}$ are the
corresponding chemical potentials of linkers. First, we consider
the situation where only linkers of type $AB$ are present (this
corresponds to the experimental conditions of
Refs.~\cite{Mirkin00,Mirkin03}). For this system, the grand
canonical partition function is:

\begin{eqnarray}
Z=\sum_{N_A, N_B, n} \, g(N_A, N_B, n) \, Q_{AB}^n \,\,
e^{  \frac{\mu^c_{A} N_A+\mu^c_{B} N_B}{T} }, \label{pf1}
\end{eqnarray}
where $N_A$ and $N_B$ are the number of colloids $A$ and $B$,
respectively, $n$ is the number of nearest neighbor $AB$ colloidal
pairs for a given realization of the grand canonical ensemble, and
$g(N_A,N_B,n)$ is the total number of possible configurations with
$n$ nearest neighbor $AB$ pairs for a given values of $N_A$ and
$N_B$. The sum with respect to $n$ extends over all values of $n$
consistent with the fact that there are $N_A$ and $N_B$ colloids
$A$ and $B$ present. The number of links between a given pair of
colloids depends on the surface density of probe strands on the
$A$ and $B$ colloids. If there is only one pair of probe strands
per contact, then this pair can accommodate at  most one linker
and the bound linkers obey Fermi-Dirac statistics.  In contrast,
when there are (infinitely) many probe strands per contact, then
the binding of linkers is determined by Bose-Einstein
statistics~\cite{NoteInt}. In general, the maximum number of
linkers per contact is a number $M\ge 0$. For $\infty> M > 1$, the
bound linkers obey fractional statistics. In practice, the local
coverage of probe molecules fluctuates~\cite{Mirkin03}. We assume
that the maximum number of linkers per bond, obeys Poisson
statistics. Moreover, we assume that the values of $M$ for
different colloid pairs, are uncorrelated.  If we average over all
possible values of $M$, we obtain the following expression for the
grand partition function of linkers (for a given $AB$ colloidal
pair):
\begin{eqnarray}
Q_{AB}=\sum_{M=0}^\infty \, p(M)\, \,\sum_{L=0}^M e^{
\frac{L(\mu_{AB}-\epsilon_{AB})}{T} }
=\frac{1-z\,e^{\overline{M}(z-1)}}{1-z}, \label{pflinkers}
\end{eqnarray}
where  $\overline{M}$ is the average value of $M$, and $z \equiv
e^{ \frac{\mu_{AB}-\epsilon_{AB}}{T} }$. $\epsilon_{\alpha
\beta}(T)$ is the temperature-dependent binding free energy of a
double-stranded DNA (dsDNA) molecule connecting a given pair of
$\alpha \beta$ colloidal particles. By tuning $\overline{M}$, we
change the effective fractional statistics of the linkers. The
Bose-Einstein limit ($Q_{AB}=1/(1-z)$) is recovered when
$\overline{M} \gg 1$. The Fermi-Dirac limit is only recovered when
$M$ is fixed, and equal to one. The physics of DNA melting and its
effect on the phase behavior of the DNA-colloid system is
described exclusively by $\epsilon_{\alpha \beta}$.

It is straightforward to generalize the model  (Eq. (\ref{pf1})),
to the case where all three possible types of linkers are present.
This results in  multiplicative factors in Eq. (\ref{pf1}) that
account for all possible configurations of all linkers:
$Q_{AB}^{n_{AB}}\,Q_{AA}^{n_{AA}}\,Q_{BB}^{n_{BB}}$, where $Q_{AA}$ and $Q_{BB}$ are
defined analogously to $Q_{AB}$, Eq. (\ref{pflinkers}).

The expression given by  Eqs. (\ref{pf1}, \ref{pflinkers}) can be
interpreted as the partition function of a three-state spin model
with Hamiltonian
\begin{eqnarray}
H=\frac{1}{2}\sum_{<ij>}\hat{\sigma}_i \hat{J} \, \hat{\sigma}_j -
\sum_{i} \hat{\eta} \, \hat{\sigma}_i  \; . \label{spin1}
\end{eqnarray}
$\hat{\sigma}_i=(1,0)$, $(0,1)$, or $(0,0)$ if $A$, $B$, or a
vacancy (solvent) occupies site $i$; and
$\hat{\eta}=(\mu^c_{A},\mu^c_{B})$. The square, symmetric
interaction matrix is given by $J^{\alpha \beta}=- T\ln
\,Q_{\alpha \beta}$. We note that the effective attraction between
the colloids induced by linkers is dominated by the entropy
associated with the number of different ways of distributing $L$
linkers over $M$ bonds: in a dense phase, there are simply more
bonds. Similar, entropic mechanisms are responsible for phase
separation in binary
hard-core mixtures \cite{Daan92}, polymer-microemulsions \cite{Porte03}, and
microemulsions \cite{Cates92,Safran00}.

\par\noindent 

The Hamiltonian for a three-component lattice gas model  (Eq.
(\ref{spin1})) has been studied
extensively~\cite{Mukamel75}. The mean-field
expression for the free energy is:
\begin{eqnarray}
f&=& T \phi_A \ln \phi_A + T \phi_B \ln \phi_B \nonumber\\
&+&T \,(1-\phi_A-\phi_B)\ln (1-\phi_A-\phi_B)\nonumber\\
&+&\frac{qJ^{AA}}{2} \phi_A^2 + \frac{qJ^{BB}}{2} \phi_B^2 + q J^{AB} \phi_A \phi_B , \label{free1}
\end{eqnarray}
where $\phi_A$ and $\phi_B$ are the average volume fractions of
$A$ and $B$ colloids. The equilibrium phase behavior of the system
in the mean-field approximation follows from the analysis of this
free energy~\cite{Mukamel75}: depending on the
strength of the interaction between the colloidal species, the
system can be in a homogeneous state, or separate into two, or
even three, coexisting phases.  It is straightforward to
generalize the model to any number of colloidal species and
corresponding linkers.

We first apply the above model to analyze the experiments of
Refs.~ \cite{Mirkin00,Mirkin03}. In those experiments,  only $AB$
linkers were present in solution, and hence $J^{AA}=J^{BB}=0$. The
free-energy difference between the helix and coil states of a
dsDNA molecule, $\epsilon_{AB}(T) \equiv  \epsilon(T)$ constitutes
an input to the interaction potential. We adopt the simple form
$\epsilon(T)=s \,(T-T_m)$, where $T_m$ is the melting temperature
of a dsDNA molecule and $s \approx 10$~\cite{fnDNAmelting}. The
dissolution  curves of the DNA-linked colloidal aggregates are
shown in Fig.~\ref{figAB}. The fraction of colloids in solution is plotted
as a function of temperature. Above a critical temperature $T_c$,
all colloids are in solution.
Consistent with the measurements of Ref.
\cite{Mirkin03}, the dissolution temperature of the aggregates
increases with increasing surface coverage of ssDNA molecules
(represented by the linker occupation number $\overline{M}$) on the colloids.
In the ``Bose-Einstein'' limit ($\overline{M}\gg 1$), $T_c$ saturates
at a value $T_c=\frac{ \mu+s }{ \ln\left[\, 1 - \exp(-8/q)
\right]+s }$, as follows from Eq. (\ref{free1}) in the case of equal initial volume
fraction of colloids $A$ and $B$ ($\phi_A=\phi_B$).

The fact that the dissolution of the DNA-linked colloidal
aggregates occurs in a narrow temperature range, makes it possible
to distinguish DNA strands that differ in only a single
base-pair~\cite{Mirkin00}. This is achieved  by selecting a
temperature (``optimum stringency temperature'') where a perfectly
matched target results in the aggregation of a high proportion of
the nanoparticles, while the same amount of ssDNA with a single
mismatch causes only a small fraction of the nano-particles to
aggregate. As can be seen from Fig.~\ref{figAB}, our model reproduces the
observed behavior: For the model calculations shown in Fig.~\ref{figAB}, the
optimum stringency temperature (indicated by the arrow) is the dissolution temperature ($T$
on the binodal) for the mismatched-DNA-linked colloids, $T=1.273$
(dashed curve) (in the units of $T_m$ for the perfectly complementary
dsDNA), and we assumed that $T_m$ for the mismatched DNA is $10 \%$
lower compared with $T_m$ of the perfectly complementary DNA. At
this temperature $100\%$ of the mismatched and only $0.1 \%$ of
the perfectly complementary DNA-linked colloids are dissolved. The
selectivity is thus extremely high.

Next, we consider the effect of the varying of salt concentration
on the melting of the DNA-linked colloids. 
Added salt has two effects: It changes the DNA melting temperature and
it modifies the electrostatic colloid-colloid and DNA-colloid interactions.
For the salt concentrations, used in the experiments of Ref. \cite{Mirkin03},
all electrostatic interactions are strongly screened \cite{LukatskyInPrep}
and hardly depend on the salt concentration. However, the salt concentration does affect the 
melting temperature of the DNA.
To estimate this effect, we make use of the
known properties of poly-electrolytes, in particular, the
theoretical prediction~\cite{Manning75} that the
melting temperature of the DNA helix-coil transition is
proportional to the  logarithm of the salt concentration. This
result has been verified  experimentally (see e.g., Ref.
\cite{Blake98}). 
A more detailed analysis of this result in the context of the present work will be presented in Ref.
\cite{LukatskyFuture}.
Here we simply use the fact, that for experimentally used salt concentrations, 
the DNA melting temperature varies with  salt
concentration as \cite{Rouzina99,Blake98}: $T_m=T_m^0\, [1+\alpha \,\text{log}\,
c \, ]$, where $T_m^0$ is the DNA melting temperature at a reference
state (with a salt concentration  $I_0=0.1$ M),
$\alpha \approx 0.05$, and $c=I/I_0$. We then obtain the following
expression for $z$: $z=\exp [ \frac{\mu+s \,(1+\alpha\,
\text{log}\, c)}{T}-s  ]$, where $\mu$ and $T$ are expressed in
the units of $T_m^0$. This expression reproduces the
experimentally observed dependence of $T_c$ on  $\sim \text{log}\,c$~\cite{Mirkin03}.

\par\noindent 
\begin{figure}[ht]
\begin{center}
\centerline{\psfig{figure=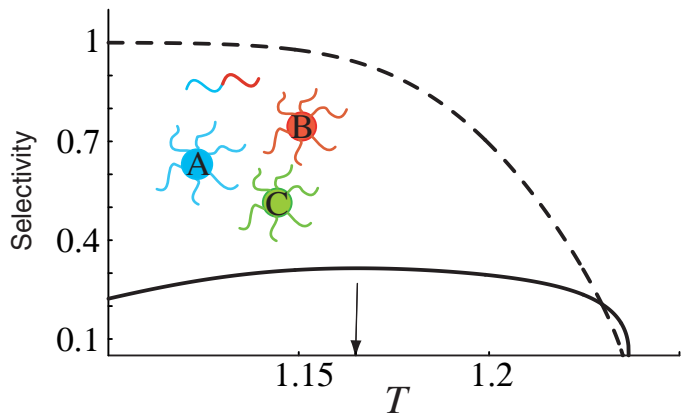,height=5.5cm,width=9cm}}
\caption{Selectivity of the phase behavior of a ternary
suspension consisting of $A$, $B$, and $C$ colloidal species and
one type of ssDNA linker that is perfectly complementary to $AB$
pair and slightly mismatched with respect to $AC$ pair (see text).
The solid (dashed) curves show that the presence of $AB$ linkers
induces a large relative concentration differences $(
\phi_B-\phi_C)/(\phi_B+\phi_C)$ of $B$ and $C$ colloidal species
in the dense (solid curve) and dilute
$-(\phi_B-\phi_C)/(\phi_B+\phi_C)$ (dashed curve) phases,
respectively. Above the demixing temperature, the concentration of
all colloidal species is the same: $\phi_A=\phi_B=\phi_C=0.17$.
The parameters used in to compute this figure are: $\mu=2.5$,
$s=10$, $q=6$, $\overline{M}=3$. $T$ and $\mu$ are expressed in
the units of $T_m$ of the linker that binds $AB$. The melting
temperature of the linker connecting $AC$ was assumed to be only
$5 \%$ lower than $T_m$. The arrow indicates the optimal
stringency temperature.}
\label{figABC}
\end{center}
\end{figure}

The model described above suggests a novel approach to DNA
screening (mutation analysis). As mentioned above, colloids
functionalized with oligonucleotides have been used as probes to
differentiate perfectly complementary targets from those with
single-base mismatches (see Ref. \cite{Mirkin00}). The drawback of
the existing method is that the thermodynamic properties of
DNA-linked nanoparticle assemblies are sensitive to the parameters
of the system, such as the grafting DNA density, the
concentrations of DNA linkers and colloids, the monodispersity of
colloids, and the pH and ionic strength of the solvent. As it is
essential to maintain precisely the same experimental conditions
when performing the measurements on different target DNA strands,
mutation-screening experiments are rather demanding.

Here we propose a method that should be relatively insensitive to
unavoidable variations in the experimental conditions. To
illustrate our approach, consider a system consisting of three
colloidal species $A$, $B$, and $C$ (see Fig.~\ref{figABC}). Species $B$ and
$C$ are grafted with ssDNA molecules that have similar sequences
and only different by a mismatch (either a single-base or
multiple-base). We assume that the target ssDNA is either
perfectly complementary to $AB$ sequence, or mutated, and thus
perfectly complementary to $AC$ sequence (the method is easily
generalized to a larger possible number of mutations).

The addition of the target ssDNA, induces a two-phase separation
in the system. Importantly, the composition of the dense aggregate
(and, by implication, the composition of the remaining solution)
depends on the nature of the target ssDNA. The characteristic
mean-field phase diagram for this case is shown in
Fig.~\ref{figABC} in terms of the concentration differences
(selectivity plot) between $B$ and $C$ colloids. The target is
perfectly complementary to $AB$. The melting temperature of the
target DNA is assumed to be some $5 \%$ higher for a perfectly
matched sequence than for a mismatched sequence. The initial
concentrations of $A$, $B$, and $C$ colloids are chosen to be
equal. Our main prediction is that the sequence of the target can
be determined simply  by monitoring the concentration
\textit{differences} of the species $B$ and $C$ in either a dense
(solid curve) or dilute (dashed curve) phase. In the case shown in
Fig.~\ref{figABC}, at the optimal stringency temperature $T\approx
1.165$ (in the units of $T_m$ of the $AB$ linker), the difference
between the concentrations of $B$ and $C$ colloids in the
aggregated phase is as large as $30 \%$. If the colloids
fluorescently labeled, even small differences in concentrations
can easily be detected. In this way, a single experiment allows us
to distinguish between the perfectly complementary and mismatched
target. We stress that the selectivity of the DNA screening can be
tuned by varying the surface coverage density $\overline{M}$. The
method should be robust to moderate variations in the experimental
conditions. It is straightforward to generalize the technique to
discriminate between a larger number of possible mutations, simply
by increasing the number of colloidal probe species.

In essence, the
method allows us to map the microscopic base-pair sequence of the
target ssDNA onto the macroscopic phase behavior of the DNA-linked
nanoparticle solution.

We are grateful to W.C. Poon, T. Schilling, A. Cacciuto and S.
Tans for helpful comments. The work of the FOM Institute is part
of the research program of FOM and is made possible by financial
support from the Netherlands organization for Scientific Research
(NWO).

\bigskip


\begin{thebibliography}{9}
\bibitem{fnmelting} In Refs. \cite{Mirkin00,Mirkin03}, the word ``melting'' is used
to describe the process of dissolution of the DNA-linked colloidal
aggregates. In the present work, we avoid the use of terms
``melting'' and ``freezing'' for this transition and use instead
``dissolution'' and ``demixing'', as we argue that it is, in fact,
more closely related to a liquid-liquid phase separation. This
phase transition should also be distinguished from the melting of
ssDNA linker molecules bound to complementary ssDNA grafted to
colloids.
\bibitem{NoteInt}
At this level of description, we ignore the interaction between
different linkers.
\bibitem{fnDNAmelting} The expression for the free energy difference that we employ ($\epsilon(T)
=s(T-T_m)$, where $s \sim N |\Delta S|$), represents the first term in the expansion of the
exact free energy, $\epsilon(T) \sim N (\Delta H -T \Delta S) $,
where $N$ is the number of base pairs in a DNA molecule, $\Delta
H$ and $\Delta S$ are, respectively, the enthalpy and entropy
difference between the helix and coil states at a given
temperature per base pair (see e.g. Ref.
\cite{Rouzina99}).
\end{thebibliography}
\end{document}